\def\lsim{\mathrel{\scriptstyle{\buildrel < \over \sim}}}

\documentclass[aps,prl,preprint,superscriptaddress]{revtex4}

\usepackage{epsfig} 

\begin{document}


\title{Collective pinning of the vortex lattice  by columnar defects 
in layered superconductors}

\author{J. P. Rodriguez}
\affiliation{Dept. of Physics and Astronomy, California State University,
Los Angeles, California 90032, USA} 

\date{\today}

\begin{abstract}
The mixed phase of layered superconductors 
with no magnetic screening
is studied through a partial duality analysis of the corresponding
frustrated $XY$ model in the presence of random columnar pins.
A small fraction of pinned vortex lines is assumed.
Thermally induced
plastic creep of the vortex lattice within
isolated layers 
results in an intermediate Bose glass phase that
exhibits weak superconductivity across layers
in the limit of weak Josephson coupling.
The correlation  volume of the vortex lattice is estimated
in the strongly-coupled Bose-glass  regime at lower temperature.
In the absence of additional point pins,
no peak effect
in the critical current density 
is predicted to occur on this basis
as a function of the Josephson coupling.
Also, the phase transition observed recently inside of the vortex-liquid phase
of high-temperature superconductors 
pierced by sparse columnar defects is
argued to be a sign of dimensional cross-over.
\end{abstract}

\maketitle

\section{Introduction}
Random material defects that are
correlated along the lines of
magnetic induction   are perhaps
the most effective mechanism known
to pin down the vortex lattice in 
high-temperature superconductors\cite{corr-pin}.
A Bose glass phase with a divergent tilt-modulus
is predicted to exist 
at relatively low fields such that each
and every flux-line can be trapped by a 
correlated material defect\cite{bose-glass}.
Experimental studies of the opposite limit,
where the number of  flux lines exceeds the number
of correlated pinning sites,
have been made recently\cite{pore}\cite{poly}\cite{nano}.
Less is known theoretically about this regime in comparison to the
former relatively low-field regime.
The vortex lattice is pinned collectively here
by the sparse columnar defects
in the low-temperature limit\cite{bose-glass}. 
A Bose glass phase that displays an infinite tilt modulus is
therefore still expected.
The collective pinning effect is naturally degraded by thermal fluctuations,
however.  
In stark contrast to the case of point pinning, 
two classes of vortex lines exist 
in the case of sparse columnar defects\cite{pore}\cite{nano}:
({\it i}) those lines of vortices that are pinned down at a columnar defect,
and ({\it ii}) those lines that are not.
An {\it interstitial}  phase
is therefore  predicted to exist theoretically
at intermediate temperatures between
the Bose glass   and the vortex liquid phases,
where only a fraction of the flux lines remain pinned down
while the remaining lines are free to wander and
thereby degrade the superconductivity\cite{IL}.

High-temperature superconductors are 
also layered materials\cite{Tink}\cite{blatter}.
Thermal fluctuations of the vortex lines in layered superconductors
are larger than in those of  isotropic materials. 
This  makes them
ideal candidates to be a host for
interstitial vortex matter.
Monte Carlo simulations of the frustrated $XY$ model have been
performed recently in order  to study  extremely type-II 
layered superconductors 
in perpendicular magnetic field
with sparse columnar   pins located at random\cite{nono}.
These simulations find 
evidence for the existence of an  interstitial liquid/glass phase
as described above.
We study this possibility  here theoretically
through a duality analysis of the same layered $XY$ model\cite{jpr00a}.
An intermediate Bose glass phase that shows weak superconductivity
across layers\cite{supersolid} exists at temperatures that lie
between the vortex-liquid phase and the Bose-glass phase in the limit
of weak Josephson coupling between layers.
We argue that the transition between the strongly-coupled
Bose glass that exists in the zero-temperature limit 
and the latter weakly coupled Bose glass
is  a cross-over by
comparison with the duality transformation of the
layered $XY$ model without frustration\cite{jpr00b}.
Numerical simulations find evidence for a sharper transition,
however,
which we suggest is an artifact of the relatively coarse model
grid that was used\cite{nono}.
We also argue for the absence of a peak effect in the critical current
density of the strongly-coupled Bose glass phase 
as a function of the Josephson coupling
if no additional point pinning is present
(cf. ref. \cite{jpr04b}).
This is based on an estimate for the Larkin correlation volume 
of the vortex lattice\cite{Tink}\cite{LO}.
Last, a vortex liquid-vortex liquid transition has 
been observed very recently
in high-temperature superconductors
with sparse columnar defects\cite{nano}.
Both the material defects
and the external  magnetic field
were oriented perpendicular 
to the copper-oxygen planes.
We shall interpret this phenomenon as a dimensional cross-over transition
that exists inside of the vortex-liquid phase of the
defective superconductor\cite{jpr02},
but that is absent in the pristine superconductor\cite{unirr}.

\section {Isolated Layer}
Consider first a stack of isolated superconducting layers 
in a perpendicular external magnetic field.
The $XY$ model over the square lattice with
uniform frustration then  provides a qualitatively
correct description of the mixed phase for each layer
in the absence of Josephson coupling,
as well as of magnetic screening. 
The neglect of the latter  is valid at perpendicular fields that
are far enough above the lower-critical field such that
\begin{equation}
a_{\rm vx} \ll \lambda_L,
\label{valid1}
\end{equation}
where $a_{\rm vx}$ is the square-root of the area per vortex in each layer,
and where $\lambda_L$ denotes the London penetration depth associated
with supercurrents that flow within each layer.
A weak Josephson coupling will be turned on in the next section.
The corresponding Boltzmann distribution is set by the
sum of energy functionals
\begin{equation}
  E_{XY}^{(2)} = - \sum_{\mu = x, y} \sum_{\vec r} J_{\mu} \,
{\rm cos} [\Delta_{\mu} \phi  - A_{\mu}]
\label{2DXY}
\end{equation}
for the superfluid kinetic energy of each layer $l$
written in terms of the superconducting phase 
$\phi (\vec r, l)$.
Here $\Delta_{\mu} \phi (\vec r) =
\phi(\vec r + a \hat\mu) - \phi(\vec r)$ and
$\vec A = (0, 2\pi f x/ a)$ make up the local supercurrent,
where $f$ denotes the concentration of vortices over the
square lattice with lattice constant $a$.
The local phase rigidity $J_{\mu} (\vec r)$
is assumed to be constant over most of the 
nearest-neighbor links $(\vec r, \vec r + a \hat\mu )$
in layer $l$,
with the exception of those links in the vicinity of the pinning sites
that are located at random.
Macroscopic phase coherence is monitored by the two-dimensional (2D) 
phase rigidity given by one over the dielectric constant of the 2D Coulomb
gas ensemble\cite{cec03} that corresponds to 
the frustrated $XY$ model (\ref{2DXY}).
Vortex/anti-vortex excitations
{\it not} associated with displacements of the zero-temperature vortex lattice
are suppressed exponentially
at temperatures far below the Kosterlitz-Thouless transition.
The  total number of vortices is conserved in such case.
This then  ultimately  leads to the result
\begin{equation}
\rho_s^{(2D)}  = J_0 [1 - ( \eta_{\rm vx}^{\prime} / \eta_{\rm sw} )]
\label{rho2d}
\end{equation}
for the 2D phase rigidity,
where $\eta_{\rm sw} = k_B T / 2\pi J_0$  
is the spin-wave component of the phase correlation exponent,
and where\cite{jpr04a}
\begin{equation}
\eta_{\rm vx}^{\prime} = \pi \Bigl\langle
\Bigl[\sum_{\vec R}^{\qquad\prime} \delta\vec u \Bigr]^2\Bigr\rangle /
N_{\rm vx} a_{\rm vx}^2,
\label{etavx1}
\end{equation}
monitors fluctuations of the
center of mass of the vortex lattice\cite{jpr01}. 
Above, 
$\delta\vec u$
is the displacement field of
each vortex with respect to its   location at zero temperature.
Also above, $N_{\rm vx}$ denotes the total number of vortices, 
while $a_{\rm vx} = a / f^{1/2}$.
Finally, $J_0$ denotes the gaussian phase rigidity\cite{cec03}.
Generalized phase auto-correlation functions
$C_l [q] = \langle  {\rm exp} [ i\sum_{\vec r} q(\vec r)
\cdot \phi (\vec r, l)] \rangle_{0}$
within  an isolated layer, $l$, can also be computed
using  the Villain approximation in the
low temperature limit\cite{villain}.
This yields the form\cite{jpr01}
\begin{equation}
C_l [q] = |C_{l}[q]|
\cdot {\rm exp} [i\sum_{\vec r} q(\vec r)  \phi_0(\vec r, l)]
\label{form}
\end{equation}
for such  autocorrelation functions,
where $\phi_0 (\vec r, l)$ represents the zero-temperature configuration
of an isolated layer.
In the ordered phase, where  $\rho_s^{(2D)} (T) > 0$,
phase correlations  are found to  decay algebraicly  like
\begin{equation}
|C_l [q]|
    = g_0^{n_+}\cdot {\rm exp}\Bigl[  \eta_{2D}
\sum_{(1,2)} q(1){\rm ln} (r_{12} / r_0)\, q(2)\Bigr]
\label{clq}
\end{equation}
at the asymptotic limit, $r_{12}\rightarrow\infty$,
with a net correlation exponent
$\eta_{2D} = k_B T/ 2\pi\rho_s^{(2D)}$.
Above, $g_0$ is equal 
to the ratio of the 2D  phase rigidity
with its value at zero temperature, 
while  $n_+$ is equal
to half the number of probes in $q (\vec r)$.
Last, $r_0$ is the natural ultra-violet scale.
Phase correlations are short-range in the disordered phase,
on the other hand,
where $\rho_s^{(2D)} (T) = 0$. 
In particular, the two-point phase auto-correlation function
probed at
$q(\vec r) = \delta_{\vec r,\vec r_1} - \delta_{\vec r,\vec r_2}$ 
retains the form (\ref{form}),
but it's magnitude decays exponentially like
\begin{equation}
|C_l (1,2)| = g_0 e^{-r_{12}/\xi_{2D}} 
\label{c12}
\end{equation}
at asymptotically large 
separations, $r_{12}\rightarrow\infty$.
Here $\xi_{2D}$ denotes the 2D phase correlation length.


We shall assume now  that the array of random columnar pins
quenches-in unbound dislocations into the triangular vortex lattice 
of each layer in isolation
at zero temperature\cite{n-s}.
Direct Monte Carlo simulations of the weakly disordered 2D $XY$ model
(\ref{2DXY}) in the Coulomb gas representation
indicate that this is indeed  the case for sufficiently
low frustration, $f$ \cite{jpr-cec04}, in which case 
substrate pinning of the 2D vortex lattice
by the model grid is sufficiently weak.
Direct Monte Carlo
simulations of the corresponding 
layered $XY$ model with sparse columnar pins
confirm the above\cite{nono}.
Also, both simulations show that the dislocations 
in the vortex lattice appear either
unbound or  bound-up into neutral pairs.  
In particular, dislocations do not line up to form low-angle
grain boundaries \cite{nono}\cite{jpr-cec04}.
This  is  consistent with the incompressible nature of the
vortex lattice in the extreme type-II limit.
The motion of the most common type of  grain boundary
requires a combination of glide and climb by the two orientations
of edge dislocations of which it is composed\cite{book}.
The total number of vortices 
is {\it not} conserved when a dislocation climbs, however.
This is energetically costly in the incompressible limit.
Grain boundaries  cannot therefore
move in or move out
from the surface
of the 2D vortex lattice at the extreme type-II limit.
Last, a direct evaluation of the 2D phase rigidity,
Eqs. (\ref{rho2d}) and (\ref{etavx1}),
in the zero-temperature limit
shows that macroscopic
phase coherence persists in the limit of a dilute concentration
of unbound dislocations\cite{jpr04a}.
In particular, the thermal fluctuation of 
quenched-in dislocations 
about their home sites
results in a vortex contribution to the phase correlation exponent $\eta_{2D}$
that  is small compared to the spin-wave contribution $\eta_{\rm sw}$
in the limit of a small number 
$N_{\rm df}$ of such dislocations  compared to the number of pinned vortices,
$N_{\rm pin}$ \cite{jpr04a}:
$\eta_{\rm vx}\lsim (N_{\rm df} / N_{\rm pin})\eta_{\rm sw}$. 
The  phase-coherent vortex lattice state\cite{jpr01} thus
survives in  the presence of a dilute concentration of dislocations
as a hexatic vortex glass state\cite{chudnovsky}.
It should melt into a phase-incoherent liquid state
at a transition temperature $T_g^{(2D)}$
that is  close to the 2D melting
temperature of the pristine vortex lattice, $k_B T_m^{(2D)}\cong J/14$,
in the present dilute limit,
$N_{\rm df}\ll N_{\rm pin}$.
The  existence of such a hexatic vortex glass state
is confirmed  by direct Monte Carlo simulation
of the 2D $XY$ model (\ref{2DXY}) in the
Coulomb gas representation\cite{jpr-cec04}.
In addition, current-voltage measurements of 2D arrays of
Josephson junctions in external magnetic field indicate
that the 2D superconducting/normal transition at $T = T_g^{(2D)}$
is second order\cite{arrays}.

\section {Josephson Coupling}
Let us   now add     a weak Josephson coupling energy
$-J_z {\rm cos} (\Delta_z \phi - A_z)$
to all of the vertical links
in between adjacent layers of the  three-dimensional (3D) $XY$ model.
Here, 
$J_z = J/\gamma^{\prime 2}$ is the perpendicular coupling
constant that we write in terms of the 2D phase rigidity 
at zero temperature, $J$, and of the model anisotropy parameter,
 $\gamma^{\prime} > 1$.
Also, $A_z = - (2\pi d / \Phi_0) B_{\parallel} x$ is the vector potential
that describes the parallel component of the magnetic induction,
$B_{\parallel}$, 
which we take to be oriented along the $y$ axis.  
The spacing between adjacent layers is denoted here by $d$.
Study of the field equation
that was derived by Bulaevskii and Clem in ref. \cite{B-C} 
for the difference of the superconducting phase across adjacent layers
indicates that the effect of magnetic
screening on the Josephson coupling can be neglected for
Josephson penetration depths, $\Lambda_0 = \gamma^{\prime} a$,
that are small compared to the London penetration depth associated with
(Josephson) supercurrents that flow across layers, 
$\lambda_L^{(\perp)} = \gamma \lambda_L$.
Here $\gamma=\Lambda_0/d$ is the physical anisotropy parameter.
The previous condition is then equivalent to the inequality
\begin{equation}
d\ll\lambda_L,
\label{valid2}
\end{equation}
which is notably independent of the anisotropy parameter.
We shall further assume that the optimum phase configuration
of an isolated layer is {\it unique}, despite the defective nature
of the ordered state.  The columnar pins are perfectly
correlated across layers, however.  This 
obviously implies that the zero-temperature configurations
for each layer in isolation are the same: 
\begin{equation}
\phi_0 (\vec r, l) = \phi_{\triangle^{\prime}} (\vec r).
\label{reg}
\end{equation}
{\it Notice then that a small fraction
of the vortex lines are pinned down by the columnar tracks in the case
that the  pinning per layer is sparse.}
The layered $XY$ model can then be effectively analyzed in
the selective high-temperature limit, $k_B T\gg J_z$,
through  a {\it partial} duality transformation.
It   leads to a dilute Coulomb gas (CG) ensemble
that describes the nature of the Josephson coupling in terms
of  dual charges that live on the vertical links
in between adjacent layers\cite{jpr00a}.
Below we review the results of this analysis.

Phase correlations across  layers 
can be computed from the quotient
\begin{equation}
  \Bigl\langle {\rm exp} \Bigl[i\sum_r p(r) \phi(r)\Bigr]\Bigr\rangle =
Z_{\rm CG}[p]/Z_{\rm CG}[0]
\label{quo}
\end{equation}
of partition functions for a layered CG ensemble\cite{jpr00a}:
\begin{equation}
  Z_{\rm CG}[p] = \sum_{\{n_{z}(r)\}} y_0^{N[n_z]}
\Pi_{l} C_l [q_l]
\cdot e^{-i\sum_r n_z A_z}.
\label{z_cg}
\end{equation}
Here the dual charge $n_z (\vec r, l)$ is an integer field
that lives on links between adjacent layers $l$ and $l+1$
located  at 2D points $\vec r$.
The ensemble is weighted
by a product
of phase auto-correlation functions
for isolated layers $l$
probed at the dual   charge  that accumulates onto
that layer:
\begin{equation}
 q_l (\vec r) = p(\vec r, l) +  n_z (\vec r, l-1) - n_z (\vec r, l).
\label{ql}
\end{equation}
It is also weighted
by a bare fugacity
$y_0$   that is
raised to the power
$N [n_z]$
equal to the total
 number of dual charges, $n_z = \pm 1$.
The fugacity approaches
$y_0 = J_z / 2 k_B T$ in the selective high-temperature regime,
$J_z \ll k_B T$,
reached at large model  anisotropy
 $\gamma^{\prime}\rightarrow\infty$.
Observe now that the phase factors of the correlation functions
(\ref{form}) cancel out in the CG ensemble (\ref{z_cg})
for probes $p$ that go directly across layers
due to the perfect registry across layers 
of the zero-temperature phase configurations (\ref{reg}).
These generalized autocorrelation functions can then
be replaced by   their magnitude $|C_l[q_l]|$
within the CG ensemble (\ref{z_cg}).

Expression (\ref{quo}) for  phase correlations across layers
can be evaluated  
perturbatively in the  decoupled vortex-liquid  phase at high temperature,
$T > T_g^{(2D)}$.
Consider, in particular,
the gauge-invariant
phase difference between any two layers, $l$ and $l^{\prime}$:
\begin{equation}
\phi_{l, l^{\prime}} (\vec r) = 
\phi (\vec r, l^{\prime}) - \phi (\vec r, l)
-(l^{\prime} - l) \cdot A_z (\vec r).
\label{phi_ll}
\end{equation}
Because of the cancellation of the zero-temperature
phase (\ref{reg}) mentioned above,
the lowest-order result for the corresponding phase auto-correlation function 
in powers of the fugacity $y_0$ 
of the dual CG ensemble (\ref{z_cg}) reads\cite{jpr00a}
\begin{equation}
\langle e^{i\phi_{l,l+n}}\rangle =
 (y_0/a^2)^n\Pi_{l^{\prime} = l}^{l+n-1}
\Biggl[\int d^2 r_{l^{\prime}}
|C_{l^{\prime}}(\vec r_{l^{\prime}})| \Biggr]
|C_{l+n} (-\Sigma_{l^{\prime} = l}^{l+n-1} \vec r_{l^{\prime}})|
\label{c_l,l+n:a}
\end{equation}
at zero parallel field, $A_z = 0$.
Here  $|C (\vec r_{12})|$
is the the magnitude (\ref{c12}) 
of the phase auto-correlation function
for an isolated layer.
At $n=1$,
the above   expression reduces to  Koshelev's formula for the inter-layer
cosine in the vortex liquid phase\cite{kosh96}:
$\langle {\rm cos}\, \phi_{l, l+1}\rangle =
y_0 \int d^2 r |C_l(\vec r)||C_{l+1}^* (\vec r)| / a^2$. 
Only short-range phase coherence  exists in the disordered phase of
isolated layers at $T > T_g^{(2D)}$
over a scale equal to  the phase correlation length,  $\xi_{2D}$.
Koshelev's formula hence yields a result  of order
$g_0^2 y_0 (\xi_{2D} / a)^2$ for the inter-layer cosine, 
or equivalently
\begin{equation}
\langle {\rm cos}\, \phi_{l, l+1}\rangle \sim
g_0^2 (J / k_B T) (\xi_{2D} / \Lambda_0)^2.
\label{cosine1}
\end{equation}
Here $\Lambda_0 = \gamma^{\prime} a$ is the Josephson penetration length. 
The last factor in expression (\ref{c_l,l+n:a})
constrains the $2n$-dimensional integral
at  larger separations between layers, $n\geq 2$. 
Its effect
can be neglected in the asymptotic large-$n$ limit, however.
This  yields the principal dependence\cite{jpr02}
\begin{equation}
\langle e^{i\phi_{l,l+n}}\rangle \propto 
 \Bigl(y_0 
\int d^2r |C (\vec r)|
 /a^2\Bigr)^n  
\label{c_l,l+n:b}
\end{equation}
for the phase autocorrelation function
at large separations between layers, $n\rightarrow\infty$.
The argument that is  raised to the power $n$ on the
right-hand side of Eq. (\ref{c_l,l+n:b}) above is hence of order
$g_0 y_0 (\xi_{2D} / a)^2$. 
The prefactor that is not shown above on the right-hand side
decays only polynomially with the layer separation $n$ 
(see refs. \cite{jpr00a} and  \cite{jpr02}).
Observe now that the phase correlation length across layers,
$\xi_{\perp}$,
is equal to the inter-layer spacing $d$ when the 
former argument is equal to $1/ e$. 
This occurs at a
dimensional  cross-over\cite{shenoy}\cite{glaz-ko00} field
%
\begin{equation}
  f \gamma_{\times}^{\prime 2} \sim g_0 (J/k_B T) (\xi_{2D}/a_{\rm vx})^2 ,
\label{2d-3d}
\end{equation}
given in units of the naive  decoupling scale $\Phi_0/\Lambda_0^2$,
that separates 2D from 3D vortex-liquid behavior \cite{jpr00a}\cite{jpr02}.
This cross-over field is traced  out in fig. \ref{phasedia}.
At  a fixed field, $f\gamma^{\prime 2}$,
we generally conclude that phase coherence
across a few to many layers is absent
in the decoupled vortex-liquid phase 
that lies at high temperature $T > T_{\times}$.
Last, observe that the perturbative result
(\ref{c_l,l+n:b}) for the phase correlation
across a macroscopic number of  layers $n$ 
diverges with the 2D phase correlation length $\xi_{2D}$
at the 2D hexatic vortex  glass transition.
This implies that a transition  to
a Bose glass occurs at a critical temperature
$T_{bg}$ that lies inside of the window $[T_g^{(2D)}, T_{\times}]$,
below which strict long-range phase coherence exists across a macroscopic
number of layers: $\xi_{\perp}\rightarrow\infty$.  
Indeed, the above  perturbative result 
for the phase auto-correlation function across layers
indicates that the Bose-glass melting  transition occurs at a field
approximtely $e$ times smaller than the 2D-3D cross-over field
(\ref{2d-3d}),
in which case the argument raised to the power $n$ on the right-hand side
of Eq. (\ref{c_l,l+n:b}) is set to unity instead.

Consider now
temperatures below the 2D hexatic vortex-glass transition,
$T < T_g^{(2D)}$,
where the 2D phase auto-correlation functions  decay algebraicly
following Eq. (\ref{clq}).
A Hubbard-Stratonovich
transformation
of the dual CG ensemble (\ref{z_cg})
reveals\cite{jpr97} that it is
equivalent  to
a  renormalized Lawrence-Doniach (LD) model
that  shows no explicit dependence on
the component of the magnetic field perpendicular to the layers:
\begin{equation}
E_{\rm LD} = 
\rho_s^{(2D)} \int d^2 r
\sum_{l} \Biggl[
{1\over 2}(\vec\nabla\theta_l)^2
-\Lambda_0^{-2}
{\rm cos}\, \theta_{l, l+1} \Biggr],
\label{e_ld}
\end{equation}
where
$\theta_{l, l+1} (\vec r)  =  
\theta_{l+1} (\vec r) - \theta_l (\vec r) - A_z (\vec r)$.
The local phase rigidity of  each layer in the LD model is equal
to the macroscopic one, $\rho_s^{(2D)} = k_B T/2\pi\eta_{2D}$,
while the Josephson coupling in the LD model 
is set by the Josephson penetration length, 
$\Lambda_0 = \gamma^{\prime} a$.
Also, the LD model inherits the ultra-violet cutoff $r_0$
from the auto-correlation functions (\ref{clq}) of isolated layers
in the ordered phase.
A standard thermodynamic analysis\cite{jpr02}
then yields that
the strength of the local Josephson coupling is
given by
\begin{equation}
\langle {\rm cos}\, \phi_{l, l+1}\rangle = y_0
+ g_0 \langle {\rm cos}\, \theta_{l, l+1} \rangle.
\label{cosine2}
\end{equation}
%
Likewise, the system shows phase rigidity across a macroscopic
number of layers equal to\cite{jpr02}
\begin{equation}
\rho_s^{\perp}/J_z = g_0 \langle {\rm cos}\, \theta_{l, l+1} \rangle.
\label{rho_perp}
\end{equation}
%
A gaussian approximation of the LD model (\ref{e_ld})
yields the result
\begin{equation}
\langle {\rm cos}\, \theta_{l, l+1} \rangle = 
(r_0/\Lambda_J)^{\eta_{2D}}
\label{cos_ld}
\end{equation}
for the LD ``cosine'',
where $\Lambda_J$ is of order $\Lambda_0$.
Notice how  the dependence of the LD model 
with the perpendicular field  enters implicitly through 
the natural ultraviolet cutoff $r_0$,
which lies somewhere in the range between the scale of the model
grid, $a$, and the inter-vortex scale, $a_{\rm vx}$.
Because we have  $g_0 (T)  = \rho_s^{(2D)} (T) / \rho_s^{(2D)} (0)$,
inspection of Eq. (\ref{rho_perp}) implies that 3D scaling of the
phase rigidities breaks down at small LD cosines,
$\langle {\rm cos}\, \theta_{l, l+1} \rangle \ll 1$.
In particular, since $g_0\leq 1$, Eq. (\ref{rho_perp})  implies that 
only  weak superconductivity can exist
across a macroscopic number of layers in such case:
$\rho_s^{\perp}\ll J_z$. 
By Eq. (\ref{cos_ld}), this requires weak enough Josephson
coupling such that the perpendicular field/anisotropy be larger than
\begin{equation}
f\gamma_{D}^{\prime 2} \sim  (r_0/a_{\rm vx})^2 e^{1/\eta_{2D}}. 
\label{decouple}
\end{equation}
The above decoupling scale is  astronomically large, however,
at low temperature
$\eta_{2D}\ll 1$.

The phase diagram displayed by
fig. \ref{phasedia} in conjunction with the physical
properties that are listed by phase in Table \ref{proper}
summarize the above  predictions.
At zero temperature, 
the correlated nature of the pinning implies that
optimum superconductivity exists across layers:
$\langle {\rm cos}\, \phi_{l, l+1}\rangle
= 1 = \rho_s^{\perp}/J_z$.
All of the vortex lines are either pinned or caged-in
by the columnar tracks in such case.
This phase is therefore a strongly-coupled Bose glass.
Quasi-2D plastic creep of the vortex lattice that is  driven
by  thermally fluctuating  edge dislocations\cite{book}
sets in at fields/anisotropies 
above the decoupling scale, $f\gamma_D^{\prime 2}$,
and this 
degrades the superconductivity across layers
in comparison to the  result expected from scaling the
2D superconductivity inside of each isolated layer\cite{jpr02}.
We shall refer to this phase as a weakly-coupled Bose glass.
Suppose now  that  the Bose-glass melting transition is continuous 
(to be argued for below),
and compare    the result  for
the inter-layer cosine at low temperatures $T < T_g^{(2D)}$, 
Eqs. (\ref{cosine2}) and (\ref{cos_ld}),  
with the result
for the same quantity
deep inside the decoupled vortex-liquid phase,
Eq.  (\ref{cosine1}) at $\xi_{2D}\sim a_{\rm vx}$.
It   suggests that the 2D correlation exponent in Eq. (\ref{cos_ld})
should be replaced by   an effective exponent of order unity
in the vicinity of the Bose-glass melting  transition
due to  the proximity of the vortex-liquid result, Eq. (\ref{cosine1}).
This indicates that   the
decoupling field/anisotropy (\ref{decouple}) is {\it not} exponentially  
big there,  unlike the low-temperature limit. 
Last, the  dual CG ensemble (\ref{z_cg}) that was used to obtain the
above  results is formally identical
to the  one derived from  
the layered $XY$ model with no frustration\cite{jpr00b},
provided that the 2D transition at $T = T_g^{(2D)}$ is second order.
The latter is consistent with current-voltage measurements of
2D arrays of Josephson junctions in
external magnetic field\cite{arrays}.
The standard 3D $XY$ model
exhibits a unique order-disorder transition, however,
despite the presence of extreme anisotropy\cite{shenoy}.
The  former equivalence  
then indicates that the transition between the weakly-coupled and
strongly-coupled Bose glass phases must be a decoupling cross-over,
and not a true phase transition. 
The unique 3D $XY$ transition 
is identified instead with the Bose-glass melting transition,
at which point long-range phase coherence across layers vanishes.
Last, study of the CG ensemble (\ref{z_cg}) indicates that
a first-order decoupling
transition\cite{daemen} occurs at temperatures outside of the 2D critical
regime,
$\xi_{2D}\sim a_{\rm vx}$, 
due to  the absence of a diverging length scale. 
[See  fig. \ref{phasedia},  and see  ref. \cite{jpr00a}, Eq. (62)].

\section{Discussion and Conclusions}

The following physical picture emerges from the previous analysis.
A dilute concentration of straight lines of edge dislocations
that are quenched in by the sparse columnar pins thread
the vortex lattice from top to bottom in the zero-temperature limit.
The pinning of the vortex lattice
is therefore collective\cite{bose-glass}.
It has a transverse scale for
positional correlations, $R_c$, that is  set by the separation between unbound
dislocations\cite{jpr04a}\cite{evetts}, 
and a longitudinal scale for positional correlations,
$L_c$, given by the length of the columnar pins.
The Larkin correlation volume\cite{Tink}\cite{LO}, $L_c R_c^2$, 
is then notably independent of
the strength of the Josephson coupling.
No peak effect in the critical current density is therefore expected
as a function of field/anisotropy,
$B_{\perp}\Lambda_0^2/\Phi_0$, in the zero-temperature limit.
It is important to observe that a macroscopic scale $L_c$ for correlations
of the vortex lattice along the (perpendicular) field direction 
is not possible in the case of point pinning for weakly
coupled layers (cf. refs. \cite{jpr04a} and \cite{kes}). 
Double-kink excitations that lie within the glide planes of the
lines of dislocations\cite{book},
as well as bound pairs of dislocations,
are excited thermally 
within isolated layers
at elevated temperatures $T\sim T_g^{(2D)}$.
In the presence of weak Josephson coupling, however,
such excitations are possible
only at temperatures above the decoupling cross-over scale, $T_D$,
where they act to degrade phase coherence across layers. 
The integrity of the thermally fluctuating lines of
dislocations is then finally
lost at the Bose-glass melting temperature $T_{bg}$
that lies above $T_D$, at which point macroscopic phase coherence
across layers also vanishes. 
Last, the nature of  phase correlations in isolated layers,
Eqs. (\ref{form})
 and (\ref{reg}),
indicates that
the dilute concentration of vortex lines that are  pinned
to columnar defects at zero temperature
remain pinned up to the temperature that marks the end
of the  2D critical regime, $\xi_{2D}\sim a_{\rm vx}$ 
(cf.  refs.  \cite{pore} and \cite{nano}).  
Both the weakly coupled Bose-glass regime and the 3D vortex-line-liquid
regime that straddle Bose-glass melting  at $T = T_{bg}$ 
lie below this cutoff in temperature
at field/anisotropy in the quasi-2D regime, $f\gamma^{\prime 2} > 1$.
They can therefore both be properly indentified as
{\it interstitial} phases\cite{IL}\cite{nono}, 
where the remaining lines of unpinned vortices 
show considerable thermal fluctuations
in the form of plastic creep.

Recent Monte Carlo simulations of the same $XY$ model studied here
for the mixed phase of layered supercondunctors with
sparse columnar   pins also find an intermediate
regime between the Bose glass
and the vortex liquid
that exhibits relatively low phase rigidity
across layers\cite{nono}.  
The present calculation strongly suggests the  identification of this
intermediate phase with the weakly coupled Bose glass 
that is shown in fig. \ref{phasedia} and that is  listed in Table \ref{proper}.
Measurements of the tilt modulus of the vortex lattice
by these Monte Carlo simulations
indicate that the boundary that separates
the strongly-coupled Bose glass 
from the intermediate phase represents a true phase transition, however,
as opposed to a crossover.
This may be an artifact of the $XY$ model grid,
which could be checked by performing simulations 
over finer model grids (or smaller $f$).
Last, the anisotropy that was used in the simulations 
reported in ref. \cite{nono}
was only moderate: $f\gamma^{\prime 2} = 1$. 
Eq. (\ref{2d-3d}) then predicts that a decoupled vortex liquid emerges
outside of the 2D critical region, 
$\xi_{2D}\sim a_{\rm vx}$, at high temperature
$k_B T\gg J$.  

The effects of sparse correlated pinning on the mixed phase of
layered superconductors have also been studied recently
in experiments  on Bismuth-based high-temperature superconductors 
that were irradiated to produce columnar tracks.
An intermediate ``nanoliquid'' phase is observed at temperatures
and perpendicular magnetic fields that lie just above the
melting line of the vortex lattice\cite{nano}.
This intermediate  vortex liquid phase shows a resistivity
ratio between  
the perpendicular field direction and the parallel layer direction
that is at least an order of magnitude smaller than that shown
by the more anisotropic vortex liquid phase that lies at higher temperature.
It is very possible then that the boundary separating the two liquid
phases observed experimentally
is just the dimensional cross-over line
(\ref{2d-3d}) shown in fig. \ref{phasedia}, 
at which point the phase correlation length across layers becomes
equal to the inter-layer spacing.
This identification requires that the
vortex lattice of   
the unirradiated crystal  {\it sublimate} into the decoupled
vortex liquid  phase, however,
since the intermediate 3D liquid
phase is absent in such case\cite{nano}.
A direct sublimation transition between a vortex solid
and a decoupled vortex-liquid phase
is in fact consistent with previous experimental studies of the
unirradiated system\cite{unirr}.
It is also predicted to occur theoretically in the vortex lattice state
of  pristine layered superconductors at sufficiently weak Josephson coupling,
provided that the vortex lattice in isolated layers melts through
a first-order transition\cite{jpr02}.
Last, the Bismuth-based high-temperature superconductor that was 
studied experimentally  in ref. \cite {nano}
is highly anisotropic\cite{blatter},
with a zero-temperature London penetration depth
of about $\lambda_L(0) = 0.2 \mu {\rm m}$,
and with  a layer spacing of $d = 1.5 {\rm nm}$.  
The first condition (\ref{valid1})
for the extreme type-II limit then yields a
threshold field of $500$ G at zero temperature, which is 
in the general vicinity of the 
observed nanoliquid phase. 
The second condition ({\ref{valid2}) for the extreme type-II limit,
on the other hand, is easily met.

A poly-crystalline vortex lattice phase is also observed
below the melting line
of the same Bismuth-based high-temperature superconductor\cite{poly}.  
As discussed previously at the end of section II,
grain boundaries do not occur in the vortex lattice 
of the frustrated $XY$ model (\ref{2DXY}) used here\cite{nono}\cite{jpr-cec04}, 
since it describes  incompressible vortex matter.
More generally however, these measurements find  that the
vortex solid 
melts  through  a second-order phase transition at perpendicular
magnetic fields above a certain critical point\cite{pore}.
This is consistent with the 3D-XY universality class 
for the Bose-glass melting  transition
that was argued for at the end of the previous section.
The same set of experiments find that  the melting transition of the vortex
lattice becomes first-order at fields below the critical point\cite{pore}.
This phenomenon is then consistent with 
the first-order decoupling transition\cite{jpr00a}\cite{daemen} 
argued for at the end of the previous section
at temperatures outside of the
2D critical regime, $\xi_{2D}\sim a_{\rm vx}$
(see fig. \ref{phasedia}).

In conclusion, an intermediate
Bose glass phase that shows weak superconductivity
across layers, $\rho_s^{\perp}\ll J_z$,
 exists at weak coupling in 
extremely type-II layered superconductors
in   external magnetic field,
with only  a  sparse arrangement of columnar pins
oriented perpendicular to the layers.
This phase is predicted to melt into a
3D vortex-line liquid that shows phase coherence across layers
on length scales  that are large compared to the spacing 
between adjacent layers.
We believe that the phase transition observed
recently inside of the vortex-liquid regime of high-temperature superconductors
pierced by sparse columnar tracks\cite{nano}
reflects  layer decoupling  by such a 
3D vortex-line liquid\cite{jpr02}.
This proposal is consistent with the absence of such a
dimensional cross-over transition in
the vortex-liquid phase of the unirradiated 
(pristine) superconductor\cite{unirr}.


\acknowledgments  The author thanks Y. Nonomura for correspondence.


\begin{figure}
\includegraphics[scale=0.36, angle=-90]{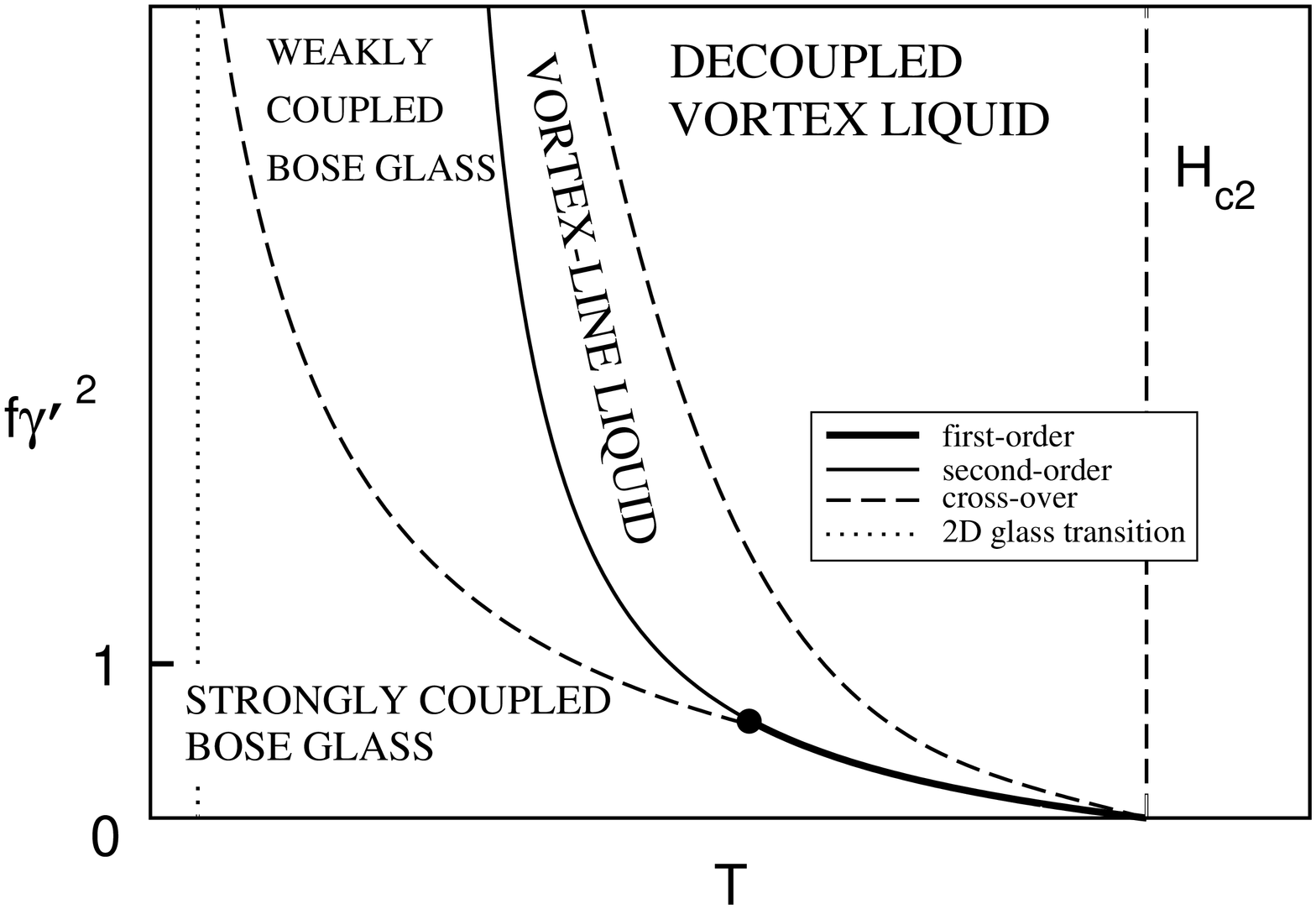}
\caption{Shown is the proposed phase diagram 
under the assumptions that 
the   2D vortex lattice 
contains a dilute concentration of unbound dislocations
and that it  melts through a
continuous phase transition (see ref. \cite{arrays}).
The dashed line inside of the Bose glass phase corresponds
to the decoupling crossover, Eq. (\ref{decouple}), while the
dashed line inside of the vortex-liquid phase corresponds
to the 2D-3D crossover, Eq. (\ref{2d-3d}).
The concentration of in-plane vortices, $f$, is held fixed,
and a   mean-field temperature dependence,
$J\propto T_{c0} - T$, is assumed. 
Also,
the effect of substrate pinning by the 2D model grid is neglected.}
\label{phasedia}
\end{figure}  

\begin{table}
\begin{center}
\begin{tabular}{|c|c|c|c|}
\hline
 regime/phase &  $\langle{\rm cos}\, \phi_{l, l+1}\rangle$
& $\rho_s^{\perp}/J_z$ &  $\xi_{\perp} / d$ \\
\hline
strongly-coupled Bose glass & unity  
& unity & $\infty$ \\
weakly-coupled Bose glass & fraction  
& fraction  & $\infty$  \\
vortex-line liquid & fraction 
& $0$  & unity, or greater \\
decoupled vortex liquid &  fraction
& $0$ & fraction \\
\hline
\end{tabular}
\caption{Listed are
the ``cosine'', the phase rigidity and the phase correlation length
across layers for the various regimes inside of  the mixed phase of
an extremely type-II superconductor
at weak Josephson coupling between layers,
and with sparse columnar pinning.}
\label{proper}
\end{center}
\end{table}

\end{document}